\documentclass[
 twocolumn,
superscriptaddress,
 amsmath,amssymb,
 aps,
prb,
raggedfooter,
]{revtex4-2}

\usepackage{graphicx}
\usepackage{dcolumn}
\usepackage{bm}
\usepackage[hidelinks]{hyperref}
\usepackage{tabularx}
\usepackage{flushend}
\usepackage{amsmath}

\hypersetup{
  colorlinks   = true, 
  urlcolor     = blue, 
  linkcolor    = blue, 
  citecolor   = red 
}

\begin{document}

\preprint{APS/123-QED}

\title{Strongly-coupled hybrid lattice-plasmons in layered cuprates}

\author{Ke-Jun Xu}
\affiliation{%
Department of Physics, University of California, Berkeley, California 94720, USA
}
\affiliation{%
Material Sciences Division, Lawrence Berkeley National Laboratory, Berkeley, California 94720, USA
}

\author{Nathan Giles-Donovan}
\affiliation{%
Department of Physics, University of California, Berkeley, California 94720, USA
}
\affiliation{%
Material Sciences Division, Lawrence Berkeley National Laboratory, Berkeley, California 94720, USA
}

\author{Stefano Agrestini}
\affiliation{%
Diamond Light Source, Harwell Campus, Didcot, Oxfordshire OX11 0DE, United Kingdom
}

\author{Jaewon Choi}
\altaffiliation{Present address: Department of Physics and Astronomy, Seoul National University, Seoul 08826, Republic of Korea}
\affiliation{%
Diamond Light Source, Harwell Campus, Didcot, Oxfordshire OX11 0DE, United Kingdom
}

\author{Mirian Garcia-Fernandez}
\affiliation{%
Diamond Light Source, Harwell Campus, Didcot, Oxfordshire OX11 0DE, United Kingdom
}

\author{Kejin Zhou}
\altaffiliation{Present address: National Synchrotron Radiation Laboratory and School of Nuclear Science and Technology, University of Science and Technology of China, Hefei 230026, China}
\affiliation{%
Diamond Light Source, Harwell Campus, Didcot, Oxfordshire OX11 0DE, United Kingdom
}

\author{Junfeng He}
\altaffiliation{Present address: Department of Physics, University of Science and Technology of China, Hefei 230026, China}
\affiliation{%
Stanford Institute for Materials and Energy Sciences, SLAC National Accelerator Laboratory, 2575 Sand Hill Road, Menlo Park, CA 94025, USA
}
\affiliation{%
Geballe Laboratory for Advanced Materials, Stanford University, Stanford, California 94305, USA
}

\author{Costel R. Rotundu}
\affiliation{%
Stanford Institute for Materials and Energy Sciences, SLAC National Accelerator Laboratory, 2575 Sand Hill Road, Menlo Park, CA 94025, USA
}
\affiliation{%
Geballe Laboratory for Advanced Materials, Stanford University, Stanford, California 94305, USA
}

\author{Young S. Lee}
\affiliation{%
Stanford Institute for Materials and Energy Sciences, SLAC National Accelerator Laboratory, 2575 Sand Hill Road, Menlo Park, CA 94025, USA
}
\affiliation{%
Geballe Laboratory for Advanced Materials, Stanford University, Stanford, California 94305, USA
}
\affiliation{%
Department of Applied Physics, Stanford University, Stanford, California 94305, USA
}

\author{Thomas P. Devereaux}
\affiliation{%
Stanford Institute for Materials and Energy Sciences, SLAC National Accelerator Laboratory, 2575 Sand Hill Road, Menlo Park, CA 94025, USA
}
\affiliation{%
Geballe Laboratory for Advanced Materials, Stanford University, Stanford, California 94305, USA 
}
\affiliation{%
Department of Materials Science and Engineering, Stanford University, Stanford, California 94305, USA 
}

\author{Zhi-Xun Shen}
\affiliation{%
Stanford Institute for Materials and Energy Sciences, SLAC National Accelerator Laboratory, 2575 Sand Hill Road, Menlo Park, CA 94025, USA
}
\affiliation{%
Geballe Laboratory for Advanced Materials, Stanford University, Stanford, California 94305, USA 
}
\affiliation{%
Department of Applied Physics, Stanford University, Stanford, California 94305, USA
}
\affiliation{%
Department of Physics, Stanford University, Stanford, California 94305, USA
}

\author{Dung-Hai Lee}
\email{dunghai@berkeley.edu}
\affiliation{%
Department of Physics, University of California, Berkeley, California 94720, USA
}
\affiliation{%
Material Sciences Division, Lawrence Berkeley National Laboratory, Berkeley, California 94720, USA
}

\author{Robert J. Birgeneau}
\email{robertjb@berkeley.edu}
\affiliation{%
Department of Physics, University of California, Berkeley, California 94720, USA
}
\affiliation{%
Material Sciences Division, Lawrence Berkeley National Laboratory, Berkeley, California 94720, USA
}

\author{Wei-Sheng Lee}
\email{leews@stanford.edu}
\affiliation{%
Stanford Institute for Materials and Energy Sciences, SLAC National Accelerator Laboratory, 2575 Sand Hill Road, Menlo Park, CA 94025, USA
}

\date{\today}

\begin{abstract}
Metallic systems with delocalized valence electrons host collective charge density oscillations known as plasmons. On the other hand, conventional insulators do not have free electrons and the low energy charge degrees of freedom are pinned to the ions. The fate of the collective charge excitations in the intermediate regime is an outstanding question. This problem is especially important for strongly correlated systems such as the layered cuprates, where unconventional superconductivity and other emergent phenomena arise from valence electrons on the border between Mott localization and itinerancy. Using resonant inelastic X-ray scattering, we track this evolution in the prototypical electron-doped cuprate Nd$_{2-x}$Ce$_x$CuO$_4$. We find a continuous transformation of the low-energy charge response: from an acoustic plasmon in the metallic regime, to a gapped hybrid mode at intermediate doping, and finally to a nearly dispersionless 139~meV excitation at half filling. Remarkably, the 139~meV excitation has approximately twice the energy of the oxygen breathing phonon responsible for the dispersion kink observed in angle-resolved photoemission spectroscopy, and is consistent with a putative 2-phonon excitation observed in Raman spectroscopy. These results establish a unified picture of collective charge excitations across the phase diagram of electron-doped cuprates, showing that such modes persist across the Mott transition via strong coupling to lattice degrees of freedom and revealing a missing link in the charge dynamics of carrier doped Mott insulators.

\end{abstract}

\maketitle


\section{Introduction}

While plasmons in good metals are well-understood~\cite{pines1}, recently there has been strong interest in the behavior of collective charge excitations in strongly correlated systems on the border between itinerancy and localization. This is because many puzzling phenomena such as strange metals~\cite{martin1,chen1,greene1,mitrano1} and high temperature superconductivity~\cite{bednorz1,keimer1} manifest in the charge sector upon doping the Mott insulating state~\cite{sachdev1,lee1,cao2}. The cuprates is an ideal model platform for investigating unconventional charge dynamics, as they host a rich phase diagram of insulating, metallic, and superconducting states~\cite{keimer1}. At half filling, the strong Coulomb potential localizes charge carriers to a single electron per site, leading to a Mott-insulating ground state. When heavily doped, these compounds behave like metals with Fermi-liquid-like properties~\cite{vignolle1,plate1}. In the intermediate doping regime, there are signatures of a strange metal phase~\cite{martin1,chen1,greene1,mitrano1}, where the charge quasiparticle is not well-defined~\cite{damascelli1}. A pressing question is how the collective charge excitations behave, if they exist at all, in the underdoped regime and across the Mott transition.

\begin{figure*}[t]
\includegraphics[width=0.9\textwidth]{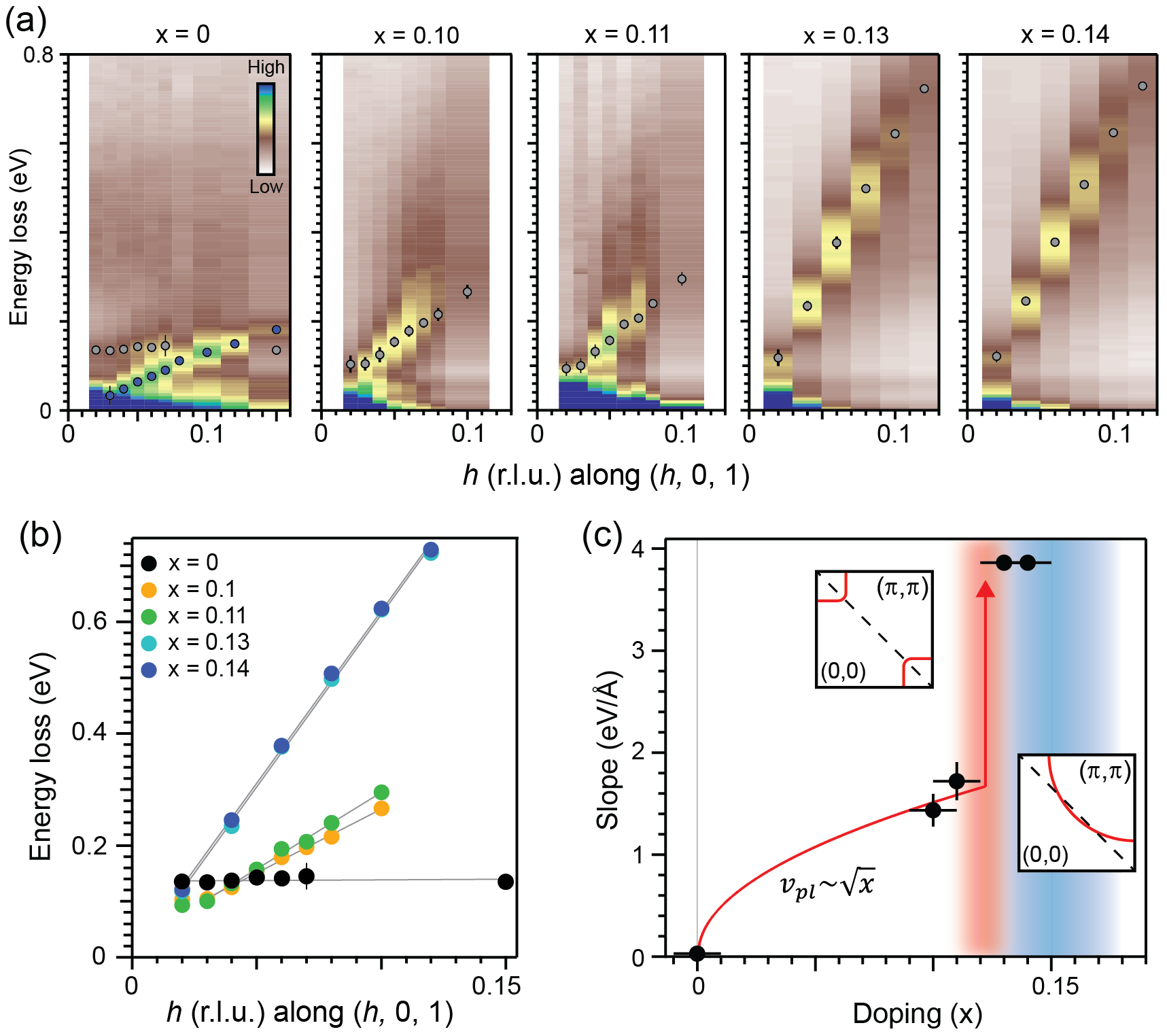}
\caption{Collective charge excitations in Nd$_{2-x}$Ce$_x$CuO$_4$ revealed by RIXS. (a) Color plots of energy-momentum RIXS spectra at different dopings measured with $\sigma$-polarized incident photons along the ($h$, 0, 1) direction. The doping levels, defined by the Ce content, are indicated above each panel. Grey dots indicate the fitted positions of the dispersive low energy charge excitations. In the x = 0 sample, blue dots represent the magnon excitation and the grey dots represent the nearly dispersionless charge excitation. (b) Dispersions of the low energy charge excitation extracted from fits of the RIXS spectra for different dopings. Uncertainties are derived from the one-sigma fit confidence intervals. Lines are linear fits to the data. (c) Slopes of the dispersions extracted from the linear fit in (b). The error bars in the slope are from the linear fit uncertainties. Red region indicates the doping where the antiferromagnetic Fermi surface reconstruction occurs. Red curve is a guide to the eye for the expected $\sqrt{x}$ dependence of the acoustic plasmon velocity at non-zero $l$. Red arrow highlights the plasmon velocity jump due to the carrier density change associated with the Fermi surface reconstruction (red shaded region). Insets in (c) show schematics of the Fermi surface with (left) and without (right) Fermi surface reconstruction in 1/4 of a Brillouin zone. Black dashed lines in the insets indicate the antiferromagnetic zone boundary. Blue color regions in (c) indicate the superconducting doping range. All spectra taken using the photon energy at the peak of the X-ray absorption spectra (see Extended Fig. 1) and at a temperature of 20 K.}
\end{figure*}

Previous resonant inelastic X-ray scattering (RIXS) measurements at the Cu $L$-edge observed sharp acoustic plasmons in electron-doped cuprates near optimal doping~\cite{hepting1}. The plasmon dispersion appears to be gapless at the Brillouin zone center for non-zero perpendicular wavevectors, consistent with a two-dimensional confinement of the electrons with weak interlayer hopping. When the interlayer hopping is strong, as in the case of the infinite layer electron-doped cuprates, there can be a finite Brillouin zone center gap induced by interlayer hopping~\cite{hepting2}. P-type plasmons have also been observed by RIXS at the O $K$-edge~\cite{nag1}, as the doped holes reside on the oxygen sites. Although the p-type plasmons tend to be weaker in intensity compared to the n-type plasmons~\cite{nag2}.

\begin{figure*}[t]
\includegraphics[width=1\textwidth]{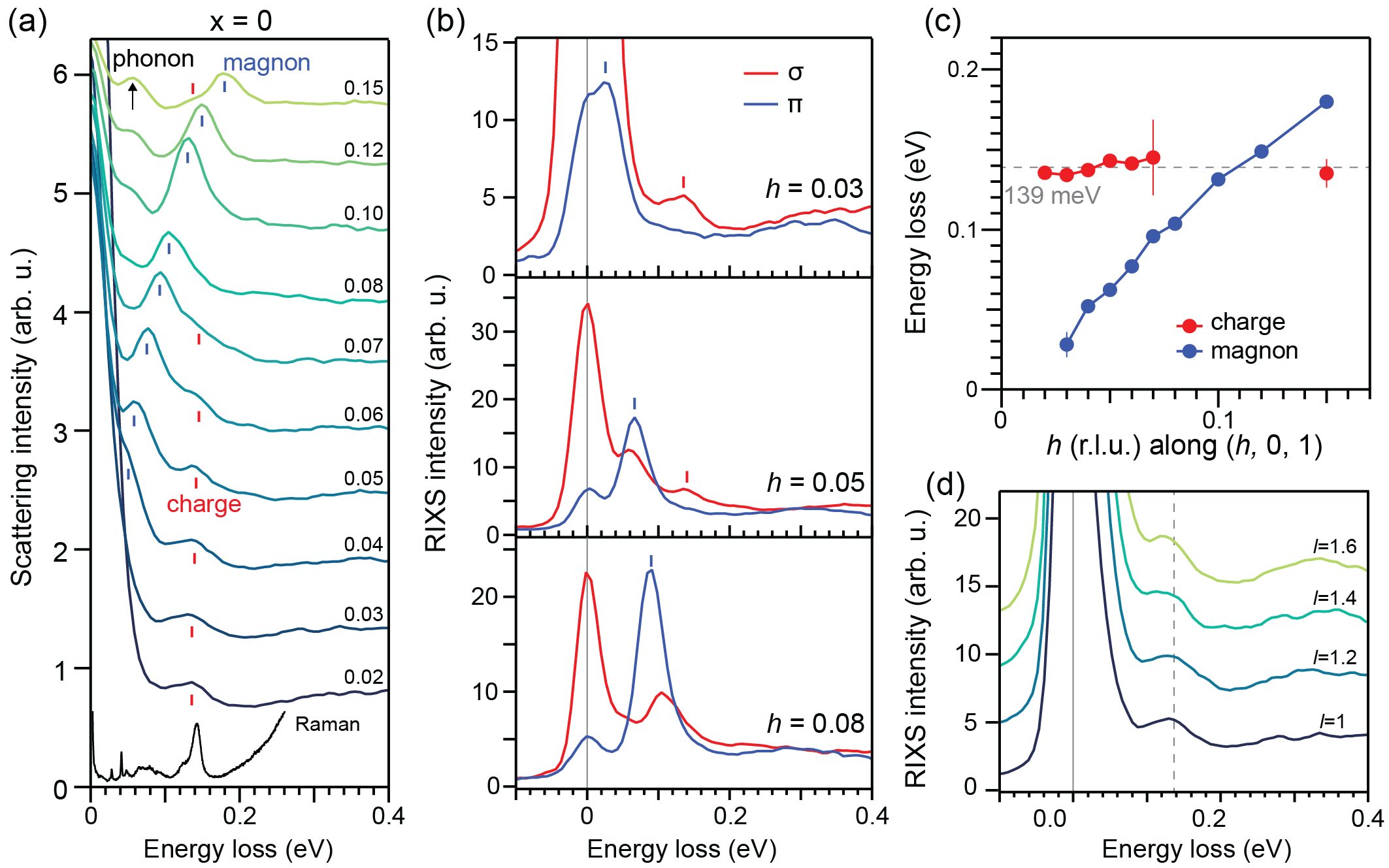}
\caption{Low energy charge excitation in Nd$_2$CuO$_4$. (a) RIXS energy distribution curves along ($h$, 0, 1) for Nd$_2$CuO$_4$. Numbers next to each curve indicate the $h$ value in r.l.u. Red markers indicate the charge excitation and blue markers highlight the magnon. Arrow pointing at a low energy single phonon peak around 50 meV. Note that this should not be interpreted as the single-phonon ($\Omega$) mode associated with the 139 meV two-phonon (2$\Omega$) excitation. Black curve at the bottom is the Raman spectroscopy data on Nd$_2$CuO$_4$. (b) RIXS curves at the indicated $h$ value along ($h$, 0, 1), for $\sigma$ and $\pi$ photon polarizations. (c) Dispersion of the magnetic and charge excitations extracted from fits of the RIXS spectra. Vertical error bars are from the one-sigma fit confidence interval. (d) RIXS spectra along (0.03, 0, $l$), showing negligible dispersion in the out-of-plane $l$ axis. Grey dashed lines in (c) and (d) indicates 139~meV.}
\end{figure*}

The relationship between these plasmons and high-transition-temperature superconductivity is a long standing question~\cite{Kresin1,ishii1,bill1}. To gain insight into this question, it is imperative to reveal how these plasmons evolve towards the lightly doped regime, where the pairing interaction is thought to be strong~\cite{emery1}. The previous RIXS works on the electron-doped cuprates mainly focused on La$_{2-x}$Ce$_x$CuO$_4$ films~\cite{lin1}, which tend to be metallic even in the nominally underdoped regime. They also cannot be synthesized as a single phase for doping below about 10\%~\cite{naito1}, thus limiting the accessible doping range. A previous work studied the doping dependence of plasmons in hole-doped La$_{2-p}$Sr$_{p}$CuO$_{4}$ and apparently found a reduction of the plasmon energy with decreasing doping down to p = 0.05~\cite{hepting3}. However, the broadness of the p-type plasmons and limited momenta points prevented an accurate extraction of the plasmon dispersions.

\begin{figure*}[t]
\includegraphics[width=0.9\textwidth]{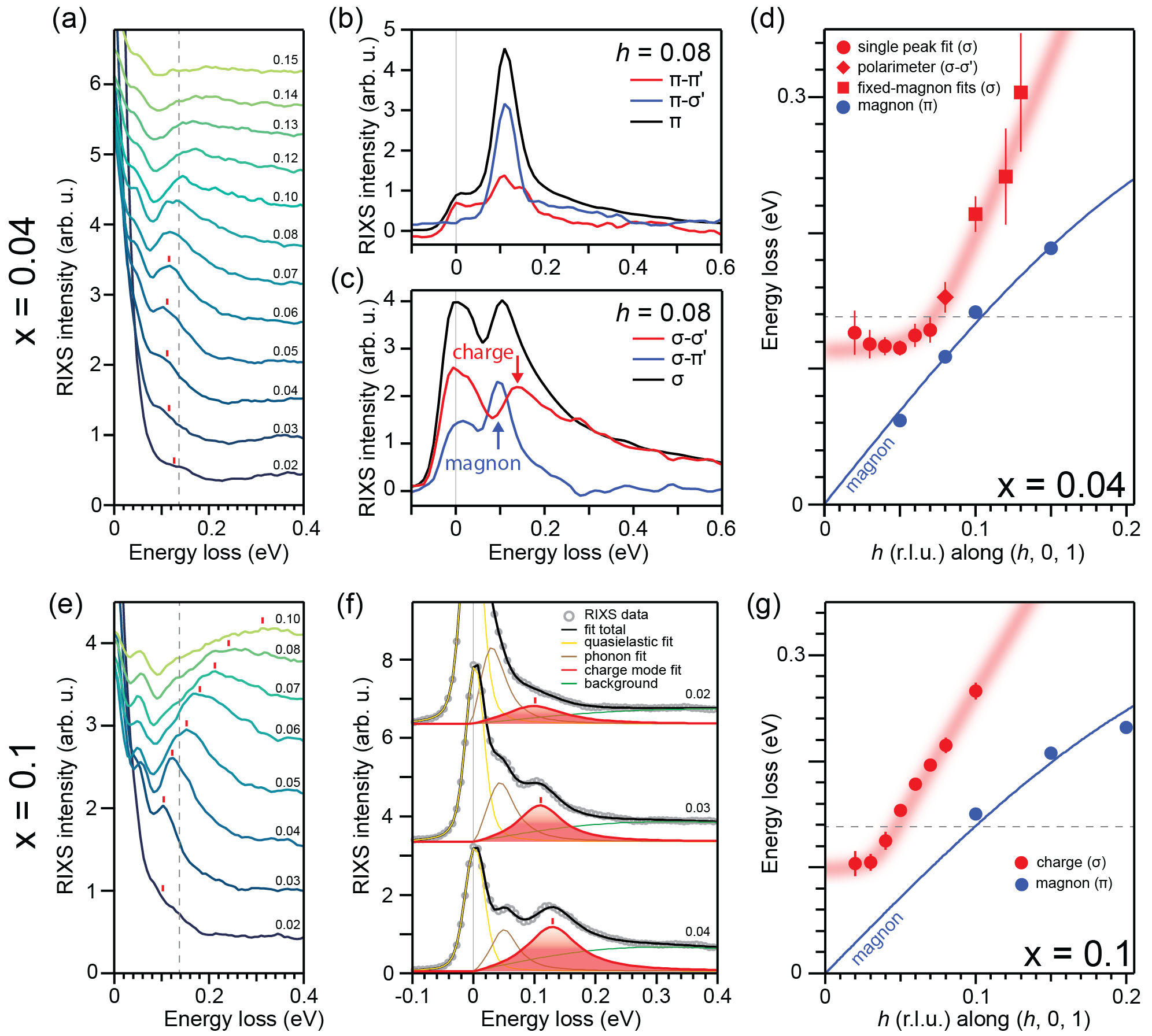}
\caption{Charge excitation dispersion anomalies in doped NCCO. (a) RIXS energy distribution curves along ($h$, 0, 1) for Nd$_{1.96}$Ce$_{0.04}$CuO$_4$. (b-c) Polarization-resolved RIXS spectra at $h$ = 0.08 for $\pi$ incident photons (b) and $\sigma$ incident photons (c). Colored curves represent polarization-resolved spectra. Black curves represent spectra with integrated outgoing polarization. (d) Dispersions of excitations in the x = 0.04 sample ($h$, 0, 1) direction. The solid red circles, solid red diamonds, and solid red squares represent the charge excitation dispersion obtained using different analysis methods (see main text and Extended data Fig. 3). Solid blue circles represent the magnon dispersion. (e) RIXS energy distribution curves for Nd$_{1.9}$Ce$_{0.1}$CuO$_4$. (f) Fitting for x = 0.1 spectra near the Brillouin zone center, with a quasielastic peak (yellow), a low energy phonon (brown), a charge excitation (red), and a background that includes bimagnon and fluorescence features (green). (g) Dispersions of excitations in the x = 0.1 sample along the ($h$, 0, 1) direction. Numbers next to each curve in (a) and (e) indicate the momentum along ($h$, 0, 1); Red markers indicate the charge excitation. Grey dashed lines in (a), (d), (e), (g) indicate 139~meV. Solid blue lines in (d), (g) are spin wave fits. Red shaded regions in (d), (g) are guides to the eye for the hybrid charge excitation (see main text and Fig. 4). Vertical error bars are from the one-sigma fit confidence interval.}
\end{figure*}

In this work, we investigate the behavior of collective charge excitations in a prototypical electron-doped cuprate Nd$_{2-x}$Ce$_x$CuO$_4$ from x = 0 to x = 0.14 using RIXS at the Cu $L_3$-edge. This doping range spans a large region of the phase diagram from the Mott-insulating parent compound to near the optimally-doped superconducting regime~\cite{krockenberger1,armitage1,xu2}. The lightly-doped regime in electron-doped cuprates is of particular interest, as recent angle-resolved photoemission spectroscopy (ARPES) studies found a hierarchy of energy gaps strongly affecting the low energy valence electronic states~\cite{xu1}: an antiferromagnetic gap that reconstructs the Fermi surface to small pockets centered at the (0, $\pi$) and equivalent points; an anomalous normal state gap that further gaps the reconstructed pockets. This cascade of energy gaps may strongly affect the low energy charge dynamics in the lightly doped regime. Our current work tracks the evolution of the collective charge excitation from the Mott insulator to the superconducting regime, revealing its transformation from a nearly dispersionless two-phonon-like mode into an acoustic plasmon through an intermediate lattice-plasmon hybrid mode.

\begin{figure*}[t]
\includegraphics[width=0.9\textwidth]{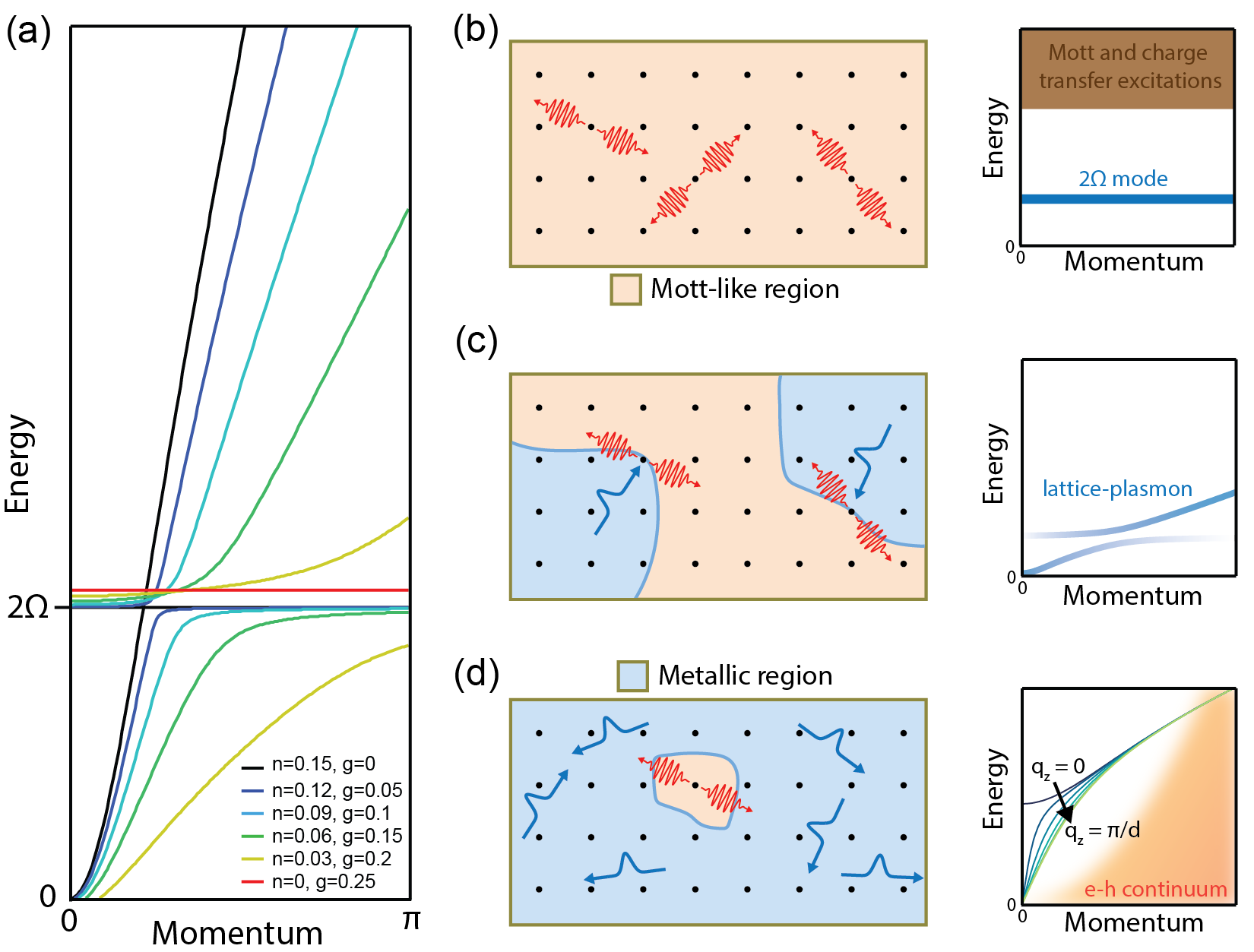}
\caption{(a) Schematic of a minimal, homogeneous effective model that captures—upon spatial averaging—the hybridization between a plasmon mode and a non-dispersive 2$\Omega$ lattice excitation (see Methods for details). The plasmon velocity decreases with decreasing doping, reflecting the reduced volume fraction of metallic regions and the emergence of a nearly non-dispersive 2$\Omega$ mode associated with the Mott insulating phase. In addition, hybridization between these modes leads to an avoided crossing. Here $n$ denotes the carrier density, $g$ the coupling constant, and 2$\Omega$ corresponds to twice the energy of the oxygen phonon mode responsible for the dispersion kink observed in ARPES. For simplicity, here we ignore the effects of antiferromagnetic Fermi surface reconstruction. (b) At half filling, one electron is localized on each Cu site by strong Coulomb interactions, producing a Mott insulating state. The low-energy charge response contains a non-dispersive 2$\Omega$ lattice excitation (right panel). Not shown are single-phonon excitations at lower energy. (c) With light doping, the system may contain nanoscale metallic regions embedded in a Mott-like matrix. Short-range plasmonic fluctuations in the metallic regions couple to nearby 2$\Omega$ lattice excitations, producing a broadened hybrid response with mixed lattice and plasmonic character (right panel). (d) In heavily doped systems with abundant metallic carriers, the collective charge excitations evolve into acoustic plasmons. The dispersion as a function of out-of-plane momentum $q_z$ is shown on the right. Here ddenotes the interlayer spacing between CuO$_2$ planes.}
\end{figure*}

\section{Collective charge excitations across the Mott transition}

We first examine the overall behavior of the low energy collective charge excitations as a function of doping. Fig. 1(a) shows energy-momentum spectra along the (h, 0, 1) direction in reciprocal space, taken with a photon energy at the peak of the Cu $L_3$-edge resonance (Supplemental Materials Fig. S1~\cite{supplemental1}). We use $\sigma$ polarized photons (E field within the CuO plane) to highlight the charge excitation more prominently. For x = 0.14, we observe dispersive acoustic plasmons, similar to those reported in previous RIXS studies~\cite{hepting1}. With decreasing electron doping, the dispersion velocity decreases, accompanied by the development of an excitation gap near the Brillouin zone center (Fig. 1a). At x = 0, in addition to a strong magnon dispersion due to the long-range antiferromagnetic order~\cite{wslee1}, there is a collective mode at around 139~meV with no observable dispersion. 

To gain further insight, we first extract an approximate dispersion velocity by linear fits at different dopings (Fig. 1(b)). Overall, we find that the plasmon velocity does not follow the expected $\sqrt x$ dependence~\cite{sarma1} at $l$ = 1 for x $>$ 0.12. The velocity exhibits a jump near x = 0.12 (Fig. 1(c)), which is associated with the antiferromagnetic Fermi surface reconstruction~\cite{dagan1} where the large Fermi surface is folded across the antiferromagnetic zone boundary to form small reconstructed pockets. Concomitantly, there is a change in the carrier density from being proportional to x at low dopings to being proportional to 1-x at high dopings. Example data fitting are presented in Supplementary Materials Fig. S2~\cite{supplemental1}. 

To clarify the evolution of the collective charge excitation, we first examine their nature at the limit of the Mott insulator. The RIXS energy distribution curves in Fig. 2(a) reveal the evolution of the low energy features in undoped Nd$_2$CuO$_4$ along the ($h$, 0, 1) direction. Apart from the elastic line, we observe a one-phonon feature at around 50~meV, a dispersing magnon feature, and an excitation near 139~meV with apparently no observable dispersion. Here, because of long-ranged antiferromagnetic order in the undoped sample, we still observe a strong magnon feature even with $\sigma$ incident photons. The magnon feature is further enhanced in the spectra taken with $\pi$ photons (Fig. 2(b)). At the same time, the 50~meV phonon and the 139~meV excitation are highlighted more prominently in the spectra taken with incident photons with $\sigma$ polarization. The dispersions of the magnon and the 139~meV excitation are summarized in Fig. 2(c). We also checked the dispersion of the 139~meV mode along the $l$ direction at a fixed $h$ value and found that there is no appreciable dispersion perpendicular to the CuO$_2$ planes (Fig. 2(d)). The dependence on the incident photon polarization and the energy scale suggest a charge origin for the 139~meV mode.

With an understanding of the behavior at half filling, we now examine the evolution of the excitation spectrum upon light doping. Fig. 3(a) shows the RIXS spectra from Nd$_{1.96}$Ce$_{0.04}$CuO$_4$ along the ($h$, 0, 1) direction, taken with incident photons of $\sigma$ polarization. The signature of the nearly dispersionless mode can still be resolved near the zone center for h $<$ 0.06, but it becomes less pronounced with a slightly softened energy of about 120~meV. For higher momentum, the charge excitation intersects with the magnon to form a single peak, as electron-doping weakens the magnon intensity such that both excitations have similar intensities. To distinguish between these two components, we perform polarization-resolved RIXS at $h$ = 0.08 (Fig. 3(b-c)). The magnon peak at around 100 meV, which manifest prominently in the cross-polarization $\pi\sigma'$ and $\sigma\pi'$ channels, possesses a lower energy than the charge peak (around 150~meV) extracted from the parallel polarization channels ($\pi\pi'$ and $\sigma\sigma'$). Furthermore, since the RIXS spectra taken with $\pi$ polarized photons is dominated by the magnon peak (e.g. Fig. 3(b)), it can be used to constrain the magnon peak position and extract the component of charge excitations in the polarization-integrate spectra taken with $\sigma$ incident polarization for h $>$ 0.06 (Supplementary Materials Fig. S3~\cite{supplemental1}).  The extracted dispersions of the charge and magnetic excitations in Nd$_{1.96}$Ce$_{0.04}$CuO$_4$ are summarized in Fig. 3(d).  As can be seen, the charge excitation becomes dispersive at h $>$ 0.07. Importantly, it appears to emerge from the dispersion-less mode near the Brillouin zone center. 

Fig. 3(e-g) shows the RIXS spectra taken with $\sigma$ polarized photons in a moderately doped sample (x = 0.1). Here, the magnon is much weaker and is not clearly observable in the spectra taken with $\sigma$ photons. Therefore, the dispersion of the charge excitations can be reliably extracted. As shown in Fig. 3(g), the collective charge excitations are well-separated from the magnon dispersion, which is extracted from the spectra taken with $\pi$ polarized photons. Interestingly, the dispersionless region near the zone center is significantly reduced compared to the x = 0.04 sample. The dispersion anomaly manifests as a bend at low momenta near (0.02, 0, 1) with an energy scale of approximately 100~meV, as is also evident in the raw data shown in Fig. 3(f). 

\section{Hybrid lattice-plasmon excitation}

The experimental observations here show that there is a complex evolution of the collective charge excitations between the undoped Mott insulator and the heavily doped metal. While the acoustic plasmons at high dopings are well-understood, the nature of the charge excitations in the undoped and lightly doped regimes are unclear. The non-dispersive mode observed in the undoped sample has an energy that is incompatible with the known phonons in the system~\cite{hofman1}. There are also no features close to 139~meV in the optical conductivity of Nd$_2$CuO$_4$~\cite{onose1}. However, this excitation energy can be associated with a lattice excitation involving two phonons, as there are single phonons with an energy near 70~meV~\cite{hofman1}. This is consistent with observations in Raman scattering~\cite{sugai1,mortiz1}, where a peak at 1178~cm$^{-1}$ (146 meV) was putatively attributed to a two-phonon overtone of the intralayer oxygen full breathing mode. Nominally, the full breathing mode is Raman-inactive, however it can become observable in Raman scattering due to lattice anharmonicity. Alternatively, this mode could be understood as a local bound two-phonon excitation~\cite{cohen1}, which may be more consistent with the unusually high intensity of the observed mode. For generality, we use the term “2$\Omega$ excitation” to phenomenologically denote an excitation with twice the single-phonon energy, without distinguishing between the weakly-coupled two-phonon overtone excitation and the strongly-coupled bound biphonon. We also note that our Raman measurements reveal a shoulder feature at around 125~meV in addition to the main peak at around 140 meV (Fig. 2(a)), indicating that multiple 2$\Omega$ lattice excitations could be relevant. 

Given the similar energy scales, we assign the non-dispersive mode in the RIXS spectra at zero doping to the same 2$\Omega$ excitation as observed in the previous Raman scattering results. We note that Raman scattering has also observed unusually intense 2$\Omega$ excitation in insulating p-type cuprates (La$_2$CuO$_4$, YBa$_2$Cu$_3$O$_{6.05}$, Bi$_2$Sr$_2$YCu$_2$O$_{8+\delta}$) at nearly the same energy as NCCO~\cite{chelwani1}. This supports the assignment of this 2$\Omega$ excitation to the same O motions within the CuO$_2$ plane, as the out-of-plane O are quite different between NCCO and the p-type parent compounds. Furthermore, La$_2$CuO$_4$ is close to being stoichiometric after annealing, unlike the other families. Thus, the observation of similar 2$\Omega$ excitations suggests that O defects likely do not play a significant role in the origin of this mode.

Considering the smooth evolution of the charge excitation between the 2$\Omega$ excitation at low dopings to the acoustic plasmon at high dopings (apart from the dispersion slope jump due to the Fermi surface reconstruction), the charge excitation at low dopings can be naturally understood as a hybrid mode formed between the plasmon and the 2$\Omega$ excitation. Fig. 4(a) shows a homogeneous effective model that captures the eigenstates of a lattice-plasmon hybrid mode. Here, the level repulsion between the hybridized branches of the lattice-plasmon effectively opens a gap at the Brillouin zone center. In the RIXS measurements, the lower branch is difficult to observe within the experimental resolution due to the presence of additional single phonon excitations and the elastic line broadening. Furthermore, if damping of the 2$\Omega$ excitation is strong, the lower branch can become broad and less visible than the upper branch (see Supplementary Materials Fig. S4~\cite{supplemental1}). We note that the excitation gap at the Brillouin zone center observed here is unlikely associated with an energy gap induced by interlayer hopping as in the case of the infinite layer electron-doped cuprates~\cite{hepting2}. In NCCO, we expect that the interlayer hopping should be reduced with decreasing doping.

Fig. 4(b-d) summarizes the overall evolution of the collective charge excitations in the electron-doped cuprates. At half filling (Fig. 4(b)), the electron wave functions are localized to the Cu ions, and the only charge excitations below the charge transfer gap of about 1.5 eV~\cite{cooper1} are single and multi-phonon excitations. With light doping, we expect that a small amount of spectral weight populates the Fermi level and acquire the potential to be mobile~\cite{shen1}. Although at low densities additional localizing interactions (e.g. Coulomb localization or disorder effects) can dominate~\cite{anderson1}, and in reality the low temperature resistivity still behaves like an insulator at low dopings~\cite{onose2}. Thus, one does not expect the presence of propagating collective charge excitations in this regime. However, it is well-known that lightly doped cuprates tend to form nanoscale phase separation~\cite{fischer1}. In this view, the lightly doped system consists of regions with incipient metallicity that support plasmon-like charge fluctuations, coexisting with Mott-like regions where the 2$\Omega$ excitation is present (Fig. 4(c)). Intra-region propagation is enabled by the plasmon component in the metallic regions, while coupling to the 2$\Omega$ excitation in the Mott regions both facilitates inter-region propagation and renormalizes the effective velocity of the mode. At high dopings, where the electron wave functions are delocalized, metallic regions dominate and the collective charge excitations behave like free plasmons (Fig. 4(d)).

\section{Implications for cuprate physics}

Our demonstration of the 2$\Omega$ excitation and its substantial coupling to plasmons is unexpected and has important implications for the nature of the cuprate ground state. Previously, hybridization between plasmons and single phonon modes has been observed in low density electron systems coupled to strongly polarizing phonons, such as graphene on SiC~\cite{koch1}. However, in the weak coupling limit, two-phonon overtone excitations arise from a high order process and are normally extremely weak compared to single phonon excitations, unlike what is observed here. Remarkably, our Raman results also show that while the single phonon peaks exhibit little change with doping, the 2$\Omega$ excitation intensity is drastically reduced with increasing doping (Supplementary Materials Fig. S5~\cite{supplemental1}). Additional experimental and theoretical efforts are required to clarify the nature of this unusual 2$\Omega$ excitation.

As the ground-state wavefunction incorporates the zero-point fluctuations of these collective modes, it is likely that they leave a measurable imprint on the unconventional phases in the cuprates, including strange metallicity and high temperature superconductivity. The observation of a similar 2$\Omega$ excitation mode in p-type cuprates with Raman scattering~\cite{chelwani1} suggests that 2$\Omega$ excitations and possibly lattice-plasmons are ubiquitous in all cuprate families. Although the hybrid mode may be difficult to resolve in the p-type cuprates with RIXS due to the relatively weaker signal. 

In the layered cuprates, even though phonons are thought not to be the progenitors of the pairing interaction, it is widely documented that they strongly interact with the electronic states~\cite{lanzara1,park1} and may play a critical role in enhancing superconductivity~\cite{he1}. In particular, anisotropic electron-phonon interactions are thought to be beneficial to d-wave superconductivity~\cite{honerkampp1}. While the coupling between single phonons and electrons is extensively studied in the cuprate context~\cite{cuk1,devereaux1}, 2$\Omega$ excitations and the associated hybrid lattice-plasmon modes have had much less considerations. How these 2$\Omega$ lattice excitations and the associated hybrid lattice-plasmon excitations interact with the antiferromagnetism is still an open question, and further work is required to elucidate any role these excitations may play in pairing. Regardless of their microscopic mechanisms and relation to electronic pairing, our results here have unambiguously demonstrated that 2$\Omega$ excitations, plasmons, and their hybrid modes play an important role in shaping the electronic landscape in the layered cuprates. 

\hspace{0.5cm}

\begin{acknowledgments}
We thank S. A. Kivelson, R. Hackl, B. Moritz, and D. Jost for insightful discussions. We are also grateful to V. R. Rocha, Y. Lyu, and J. Analytis for checking the doping levels with energy dispersive spectroscopy. The work performed at the University of California, Berkeley and Lawrence Berkeley National Laboratory was funded by the U.S. Department of Energy, Office of Science, Office of Basic Energy Sciences, Materials Sciences and Engineering Division, Contract No. DE-AC02-05-CH11231 within the Quantum Materials Program (KC2202). The work performed at Stanford and SLAC was supported by the U.S. Department of Energy (DOE), Office of Basic Energy Sciences, Division of Materials Sciences and Engineering. This work made use of the facilities at Diamond Light Source beamline I21 under proposal MM42608. 
\end{acknowledgments}

\hspace{0.5cm}

\bibliographystyle{apsrev4-2}
\bibliography{main_ref}
\raggedend
\end{document}